\documentclass[12pt,preprint,notoc]{QCEMDA}  

\usepackage{epsfig,multicol,bbm}
\usepackage{verbatim}

\newcommand\fverb{\setbox\pippobox=\hbox\bgroup\verb}
\newcommand\fverbdo{\egroup\medskip\noindent%
            \fbox{\unhbox\pippobox}\ }
\newcommand\fverbit{\egroup\item[\fbox{\unhbox\pippobox}]}
\newbox\pippobox

\title{Thermodynamics of noncommutative BTZ black hole}

\author{Rashida Jahangir$^a$ and K. Saifullah$^b$\\

$^a$Department of Mathematics, International Islamic University,
Islamabad, Pakistan \\
$^b$Department of Mathematics, Quaid-i-Azam University, Islamabad,
Pakistan  \\

Electronic address: \email{saifullah@qau.edu.pk}}

\preprint{}  

\abstract{Thermodynamics of the BTZ black hole in noncommutative
geometry is studied. We work out the Hawking temperature and entropy
which reduce to their commutative limits when the noncommutativity
parameter tends to zero. We also discuss the range of validity of
the Hawking area law in the noncommutative case and provide
graphical analysis. We see that the law is not valid unless the
outer horizon is very large.}

\begin{document}


The interest in noncommutativity was developed when it was shown
that the field theory becomes  noncommutative when matter decouples
from gravity, and the spacetime induces a noncommutative coordinate
algebra. In loop quantum gravity as well the spatial operators do
not commute and one obtains a noncommutative manifold. The
noncommutativity is introduced by means of \cite{26}

\begin{equation}
\lbrack x^{i},x^{j}]=i\theta ^{ij} ,   \label{4.1}
\end{equation}
where $\theta ^{ij}$ is antisymmetric tensor, $D$ $\times $ $D$
matrix, where $D$ is the dimension of the spacetime. It has
dimensions of (length)$^{2}$ and it discretizes the spacetime.  The
noncommutative relation given by Eq. (\ref{4.1}) induces quantum
mechanical fluctuations in the metric, $g_{\mu \nu}$. The spacetime
is quantized and the coordinates become noncommutative. In one
approach to noncommutative geometry the point-wise multiplication of
fields in the Lagrangian is replaced by the Moyal product given by
\cite{40}

\begin{equation}
(f\star g)(x)=\exp [\frac{i}{2}\theta ^{ab}\frac{\partial }{\partial
x^{a}} \frac{\partial }{\partial x^{b}}]f(x)g(x) ,   \label{4.2}
\end{equation}
where, $f$ and $g$ are functions of the coordinates. This is a
powerful method for studying field theories on noncommutative
spaces. As the parameter $\theta $ varies from zero to a positive
value we go from the commutative to the noncommutative regime.
Recently there have been some investigations on noncommutative black
holes \cite{26, 40, 24, 25}. Apart from some canonical solutions on
noncommutative spaces, such as that of the noncommutative
Schwarzschild black hole \cite {26}, most of the solutions are not
obtained directly from Einstein's field equations.


P. Nicolini \emph{et al} \cite{26} introduced noncommutativity in
the Schwarzschild black hole by taking the mass density to be a
Gaussian distribution

\begin{equation}
\rho _{\theta }(r)=\frac{M}{(4\pi \theta
)^{\frac{3}{2}}}e^{-\frac{r^{2}}{ 4\theta }} ,   \label{3.1}
\end{equation}
with minimal width $\sqrt{\theta }$, instead of the Dirac delta
function, and where the noncommutative parameter, $\theta $ is a
small positive number of the order of (Planck length)$^{2}$. Using
Eq. (\ref{3.1}) one can write the mass of the noncommutative
Schwarzschild black hole of radius $r$ in the following way
\cite{26}

\begin{equation}
m_{\theta}(r)=\int \limits_{0}^{r} 4\pi\acute{r}^{2}\rho_{\theta}
(\acute{r}) d\acute{r}=\frac{2M}{\sqrt{\pi}}\gamma
(\frac{3}{2},\frac{r^{2}}{4\theta}) ,   \label{3.2}
\end{equation}
where $M$ is mass of the commutative Schwarzschild black hole and
$\gamma ( \frac{3}{2},\frac{r^{2}}{4\theta })$ is the lower
incomplete gamma function. In Einstein's equations using the
energy-momentum tensor
\begin{equation}
(T_{\theta })_{b}^{a}=diag[-\rho _{\theta },p_{r},p^{\prime
},p^{\prime }]
 ,   \label{3.5}
\end{equation}
where $p_{r}=-\rho _{\theta }$ and $p^{\prime
}=p_{r}-\frac{r}{2}\partial _{r}\rho _{\theta }$ are pressure terms,
the noncommutative Schwarzschild solution is \cite{26}
\begin{equation}
ds^{2}=-f_{\theta }(r)dt^{2}+\frac{dr^{2}}{f_{\theta
}(r)}+r^{2}d\Omega ^{2} ,   \label{3.7}
\end{equation}
where $f_{\theta }(r)=1-\frac{4M}{r\sqrt{\pi }}\gamma
(\frac{3}{2},\frac{ r^{2}}{4\theta })$ and $d\Omega ^{2}=d\vartheta
^{2}+\sin ^{2}\vartheta d\phi ^{2}$ . It is same as if we replace
the mass term in the metric with the noncommutative mass term given
in Eq. (\ref{3.2}).


Banados \emph{et al} \cite{10} discovered a black hole solution of
Einstein's equations with a negative cosmological constant, in
($2+1$) dimensions in the commutative spacetime. The noncommutative
Banados-Teitelboim-Zanelli (BTZ) metric \cite{40, 25} was found by
using the Chern-Simons theory which is a $3$-dimensional topological
quantum field theory. In this theory the action for the field $A$ is
proportional to the integral of the Chern-Simons $3$-form i.e.
\begin{equation}
S(A)=\beta \int Tr(A \wedge dA+\frac{2}{3}A \wedge A \wedge  A) ,
\end{equation}
where $\beta =l/16\pi G$ and $G$ is Newton's constant. Classically
the system is characterized by its equations of motion which are the
extrema of the action with respect to variations of the field, $A$,
and are given by
\begin{equation}
F=dA+A \wedge  A .
\end{equation}

For the $(2+1)$-dimensional noncommutative Chern-Simons theory with
negative cosmological constant $-1/l^{2}$ the action is given by
\cite{40, 25}
\begin{equation}
\hat{S}(\hat{A}^{+},\hat{A}^{-})=\hat{S}_{+}(\hat{A}^{+})-\hat{S}_{-}(\hat{A}
^{-}) ,   \label{bt1}
\end{equation}
\begin{equation}
\hat{S}_{\pm }(\hat{A}^{\pm })=\beta \int Tr(\hat{A}^{\pm
}\stackrel{\star}{\wedge}  d\hat{A}^{\pm }+\frac{2}{3}\hat{A}^{\pm
}\stackrel{\star}{\wedge}  \hat{A}^{\pm }\stackrel{\star}{\wedge}
\hat{A}^{\pm }) .  \label{bt2}
\end{equation}
The deformed wedge product is defined as
\begin{equation}
A \stackrel{\star}{\wedge}B=A_{a}\star B_{b }dx^{a}\wedge dx^{b}
 ,   \label{bt3}
\end{equation}
where $\star $ is the Moyal product defined in Eq. (\ref{4.2}). The
noncommutative $U(1,1)\times U(1,1)$ gauge fields, $\tilde{A},$
consist of noncommutative $SU(1,1)\times SU(1,1)$ gauge fields,
$\hat{A}$ and two $U(1)$ fluxes, $\hat{B}$ \cite{40, 25}
\begin{equation}
\tilde{A}^{\pm }=\tilde{A}^{A\pm }\tau _{A}=\hat{A}^{a\pm }\tau
_{a}+\hat{B} ^{\pm }\tau _{3} .   \label{bt4}
\end{equation}
The noncommutative $SU(1,1)\times SU(1,1)$ gauge fields $\hat{A}$
are expressed in terms of the triad $\hat{e}$ and the spin
connection $\hat{w}$ as
\begin{equation}
\hat{A}^{a\pm }=\hat{w}^{a}\pm \frac{\hat{e}^{a}}{l} . \label{bt5}
\end{equation}
Using the original Seiberg-Witten map \cite{28} and the commutation
relation $[R,\phi ]=2i\theta $, where $R=r^{2}$, the solution will
be
\begin{equation}
\tilde{A}_{\mu }^{\pm }(A)=(A_{\mu }^{a\pm }-\frac{\theta }{2}B^{\pm
}\partial _{R}A_{\mu }^{a\pm })\tau _{a}+B^{\pm }\tau _{3} .
\label{bt16}
\end{equation}
The commutative gauge fields are given by \cite{25}
\begin{equation}
A^{0\pm }=\pm \frac{m(r_{+}\pm r_{-})}{l^{2}}(dt\mp ld\phi ) ,
\label{bt15a}
\end{equation}
\begin{equation}
A^{1\pm }=\pm \frac{dm}{n} ,   \label{bt15b}
\end{equation}
\begin{equation}
A^{2\pm }=-\frac{n(r_{+}\pm r_{-})}{l^{2}}(dt\mp ld\phi ) .
\label{bt15c}
\end{equation}
Using these in Eq. (\ref{bt16}), the noncommutative gauge fields can
be written as \cite{25}
\begin{equation}
\hat{A}^{0\pm }=-\frac{(m-\frac{\theta \beta }{2}m^{\prime
})(r_{+}\pm r_{-}) }{l}(d\phi \mp \frac{dt}{l}) ,   \label{bt17a}
\end{equation}
\begin{equation}
\hat{A}^{1\pm }=\pm \lbrack \frac{m^{\prime }}{n}-\frac{\theta \beta
}{2}( \frac{m^{\prime }}{n})^{\prime }]dR ,   \label{bt18}
\end{equation}
\
\begin{equation}
\hat{A}^{2\pm }=\pm \frac{(n-\frac{\theta \beta }{2}n^{\prime
})(r_{+}\pm r_{-})}{l}(d\phi \mp \frac{dt}{l}) . \label{bt19}
\end{equation}
Using Eq. (\ref{bt5}) the triad and the spin connection become
\begin{equation}
\hat{e}^{0}=(m-\frac{\theta \beta }{2}m^{\prime })(\frac{r_{+}}{l}
dt-r_{-}d\phi )+ \textrm{Order}(\theta ^{2}) ,   \label{bt20}
\end{equation}
\begin{equation}
\hat{e}^{1}=\pm \lbrack \frac{m^{\prime }}{n}-\frac{\theta \beta
}{2}(\frac{ m^{\prime }}{n})^{\prime }]dR+\textrm{ Order(}\theta
^{2}) , \label{bt21}
\end{equation}
\begin{equation}
\hat{e}^{2}=(n-\frac{\theta \beta }{2}n^{\prime })(r_{+}d\phi
-\frac{r_{-}}{l }dt)+\textrm{ Order}(\theta ^{2}) ,   \label{bt22}
\end{equation}
\begin{equation}
\hat{w}^{0}=-\frac{(m-\frac{\theta \beta }{2}m^{\prime
})}{l}(r_{+}d\phi - \frac{r_{-}}{l}dt)+\textrm{ Order}(\theta ^{2})
, \label{bt23}
\end{equation}
\begin{equation}
\hat{w}^{1}=\textrm{ Order}(\theta ^{2}) ,   \label{bt24}
\end{equation}
\begin{equation}
\hat{w}^{2}=-\frac{(n-\frac{\theta \beta }{2}n^{\prime
})}{l}(\frac{r_{+}}{l} dt-r_{-}d\phi )+\textrm{ Order(}\theta ^{2})
, \label{bt25}
\end{equation}
where prime denotes differentiation with respect to $R=r^{2}$. The
metric in the noncommutative case is given as
\begin{equation}
d\hat{s}^{2}=\eta _{ab}\hat{e}_{\mu }^{a}\star \hat{e}_{\nu
}^{b}dx^{\mu }dx^{\nu } .   \label{bt26}
\end{equation}
Since the metric coefficients are only $R$-dependent so \
$\hat{e}_{\mu }^{a}\star \hat{e}_{\nu }^{b}=$\ \ \ $\hat{e}_{\mu
}^{a}\hat{e}_{\nu }^{b} $, and we have
\begin{equation}
ds^{2}=-\hat{f}^{2}dt^{2}+\hat{N}^{-2}dr^{2}+2r^{2}\hat{N}^{\phi
}dtd\phi +(r^{2}-\frac{\theta \beta }{2})d\phi
^{2}+\textrm{Order}(\theta ^{2}) .  \label{bt27}
\end{equation}
Here

\begin{eqnarray}
\hat{f}^{2} &=&\frac{r^{2}-r_{+}^{2}-r_{-}^{2}}{l^{2}}-\frac{\theta
\beta }{2
} ,   \nonumber \\
\hat{N}^{2} &=& \frac{1}{r^{2}l^{2}}\left\{
(r^{2}-r_{+}^{2})(r^{2}-r_{-}^{2})-\frac{\theta \beta }{2}
(2r^{2}-r_{+}^{2}-r_{-}^{2})\right\}  ,  \nonumber \\
\hat{N}^{\phi } &=&\frac{-r_{+}r_{-}}{lr^{2}} , \label{bt2g}
\end{eqnarray}
where $r_{+}$ and $r_{-}$ are the outer and inner horizons of the
commutative BTZ black hole \cite{10}
\begin{equation}
r_{\pm }^{2}= \frac{l^{2} M}{2}\left\{ 1\pm \left( 1-\left(
\frac{J}{Ml} \right) ^{2}\right) ^{\frac{1}{2}}\right\} .
\label{bt28}
\end{equation}
The apparent horizons, for the noncommutative black hole, denoted by
$\hat{r}_{\pm }$ can be determined by $\hat{ N}^{2}=0$
\begin{equation}
\hat{r}_{\pm }^{2}=r_{\pm }^{2}+\frac{\beta \theta }{2}+
\textrm{Order}(\theta ^{2})  .   \label{bt34}
\end{equation}
This shows that the event horizons in the noncommutative case, are
shifted through constant $\beta \theta /2$. Both the inner and outer
horizons are equally shifted. In the limit $\theta \rightarrow 0$,
these reduce to the event horizons of the commutative case.


Let us rewrite the metric of noncommutative BTZ black hole as
\begin{equation}
ds^{2}=-fdt^{2}+g^{-1}dr^{2}+(r^{2}-\frac{\beta \theta }{2})d\chi
^{2} ,   \label{bt32}
\end{equation}
where
\begin{eqnarray*}
f=(\frac{r^{2}-Ml^{2}}{l^{2}}+\frac{J^{2}}{4(r^{2}-\frac{\beta
\theta }{2})}-\frac{\beta \theta }{2}) ,\\
g=\frac{r^{2}-Ml^{2}}{l^{2}}+\frac{J^{2}}{ 4r^{2}}-\frac{\beta
\theta
}{2}(\frac{2}{l^{2}}-\frac{M}{r^{2}}) , \\
d\chi =d\phi -\frac{J}{2(r^{2}-\frac{\beta \theta }{2})}dt .
\end{eqnarray*}

There is a coordinate singularity at $g(\hat{r}_{+})=0.$ This
singularity is removed by the use of the Painleve-type coordinate
transformation \cite{50}

\[
dt\rightarrow dt-\sqrt{(\frac{1-g}{f g})}dr .
\]
Using this transformation in metric$\ \left( \ref{bt32}\right) $ we
get

\begin{equation}
ds^{2}=-fdt^{2}+dr^{2}+2f\sqrt{(\frac{1-g}{f g})}drdt+(r^{2}-\frac{
\beta \theta }{2})d\chi ^{2} . \label{r12}
\end{equation}
Near the outer horizon we expand the functions $f$ and $g$ using
Taylor's series

\begin{equation}
f(\hat{r}_{+})=f^{\prime }(\hat{r}_{+})(r-\hat{r}_{+})+O((r-\hat{r}
_{+})^{2}\ ,  \label{r13}
\end{equation}
\begin{equation}
g(\hat{r}_{+})=g^{\prime
}(\hat{r}_{+})(r-\hat{r}_{+})+O((r-\hat{r}_{+})^{2}  . \label{r14}
\end{equation}
The surface gravity at the outer horizon is given by
\begin{eqnarray}
\kappa &=&\left. \Gamma _{00}^{0}\right\vert _{r=r_{+}} ,
\nonumber \\
&=&\frac{1}{2}\left. \left[
\sqrt{\frac{1-g}{fg}}g\frac{df}{dr}\right] \right\vert _{r=r_{+}} .
\label{3}
\end{eqnarray}
In our case this becomes

\[
\kappa =\frac{1}{2}(\sqrt{f^{\prime }(\hat{r}_{+})g^{\prime
}(\hat{r}_{+})})  .
\]
The Hawking temperature, $T=\hbar \kappa /2\pi$, at the outer
horizon takes the form

\begin{equation}
T_{h}=\frac{1}{4\pi} \hbar \sqrt{f^{\prime} (\hat r_{+})g^{\prime}
(\hat r_{+})} . \label{r16}
\end{equation}
The mass $M$ of the noncommutative black hole is given by

\begin{equation}
M=\left. \frac{r^{2}}{(r^{2}-\frac{\beta \theta
}{2})}(\frac{r^{2}}{l^{2}}+\frac{ J^{2}}{4r^{2}}-\frac{\beta \theta
}{l^{2}}) \right\vert _{r=\hat{r}_{+}} . \label{r19}
\end{equation}
If we take limit $\theta \rightarrow 0$ we recover the mass of the
commutative BTZ black hole. Thus the temperature is obtained as

\begin{equation}
T=\left. \frac{\hslash }{4\pi
}\sqrt{(\frac{2r}{l^{2}}-\frac{J^{2}r}{ 2(r^{2}-\frac{\beta \theta
}{2})^{2}})(\frac{2r}{l^{2}}-\frac{J^{2}}{2r^{3}}- \frac{\beta
\theta }{r^{3}}(\frac{r^{2}}{(r^{2}-\frac{\beta \theta }{2})}(
\frac{r^{2}}{l^{2}}+\frac{J^{2}}{4r^{2}}-\frac{\beta \theta
}{l^{2}})))} \right\vert _{r=\hat{r}_{+}} .   \label{r20}
\end{equation}
Again note that in the limit $\theta \rightarrow 0$ this reduces to
the temperature of the commutative BTZ black hole \cite{9}. The
first law of black hole thermodynamics is given by

\begin{equation}
dS=\frac{dM}{T}-\frac{\Omega }{T}dJ .   \label{1q}
\end{equation}
At the outer horizon $r=\hat{r}_{+}$ we have

\begin{equation}
M=M(\hat{r}_{+},J) ,   \label{r22}
\end{equation}

\begin{equation}
dM=\frac{\partial M}{\partial
\hat{r}_{+}}d\hat{r}_{+}+\frac{\partial M}{
\partial J}dJ .   \label{r23}
\end{equation}
But $\frac{\partial M}{\partial J}=\Omega $ \cite{24} so

\begin{equation}
dM=\frac{\partial M}{\partial \hat{r}_{+}}d\hat{r}_{+}+\Omega dJ .
\label{r24}
\end{equation}
Now using this in Eq. (\ref{1q}) we get

\begin{equation}
d\hat{S}=\frac{1}{T}\frac{\partial M}{\partial
\hat{r}_{+}}d\hat{r}_{+} ,   \label{r25}
\end{equation}
where $\hat{S}$ is the entropy of noncommutative BTZ black hole.
Putting Eqs. (\ref{r19}) and (\ref{r20}) in Eq. (\ref {r25}) we get

\begin{equation}
d\hat{S}=\left.\frac{4\pi }{\hbar }\frac{\left[
-\frac{2r}{(r^{2}-\frac{ \beta \theta
}{2})^{2}}(\frac{r^{4}}{l^{2}}+\frac{J^{2}}{4}-\frac{\beta \theta
r^{2}}{l^{2}})+\frac{1}{(r^{2}-\frac{\beta \theta
}{2})}(\frac{4r^{3} }{l^{2}}-\frac{2\beta \theta r}{l^{2}})\right]
}{\sqrt{(\frac{2r}{l^{2}}- \frac{J^{2}r}{2(r^{2}-\frac{\beta \theta
}{2})^{2}})(\frac{2r}{l^{2}}-\frac{ J^{2}}{2r^{3}}-\frac{\beta
\theta }{r^{3}}(\frac{r^{2}}{(r^{2}-\frac{\beta \theta
}{2})}(\frac{r^{2}}{l^{2}}+\frac{J^{2}}{4r^{2}}-\frac{\beta \theta
}{ l^{2}})))}}dr \right\vert _{r=\hat{r}_{+}} .   \label{r27}
\end{equation}
Taking $\hslash =1$ and integrating yields

\begin{equation}
\hat{S}=4\pi \int \left. \frac{\left[ -\frac{2r}{(r^{2}-\frac{\beta
\theta }{2})^{2} }(\frac{r^{4}}{l^{2}}+\frac{J^{2}}{4}-\frac{\beta
\theta r^{2}}{l^{2}})+ \frac{1}{(r^{2}-\frac{\beta \theta
}{2})}(\frac{4r^{3}}{l^{2}}-\frac{2\beta \theta r}{l^{2}})\right]
}{\sqrt{(\frac{2r}{l^{2}}-\frac{J^{2}r}{2(r^{2}- \frac{\beta \theta
}{2})^{2}})(\frac{2r}{l^{2}}-\frac{J^{2}}{2r^{3}}-\frac{ \beta
\theta }{r^{3}}(\frac{r^{2}}{(r^{2}-\frac{\beta \theta }{2})}(\frac{
r^{2}}{l^{2}}+\frac{J^{2}}{4r^{2}}-\frac{\beta \theta
}{l^{2}})))}}dr \right\vert _{r=\hat{r}_{+}} .   \label{r28}
\end{equation}
If we set $\theta =0$ in Eq. (\ref{r28}), we recover the standard
formula for the entropy of the commutative BTZ black hole \cite{9}

\begin{eqnarray}
S &=&4\pi \int \left. dr \right\vert _{r=\hat{r}_{+}} ,   \nonumber \\
S &=&4\pi r_{+} .
\end{eqnarray}
In order to integrate Eq. (\ref{r28}) to find entropy in the
noncommutative case we rewrite the numerator and the denominator of
the integrand as

\begin{equation}
\textrm{Numerator}=\frac{2r}{l^{2}}\left(
1-\frac{J^{2}l^{2}}{4r^{4}}\right) \left( 1-\frac{\beta \theta
J^{2}l^{2}}{r^{2}(4r^{4}-J^{2}l^{2})}\right) + \textrm{Order}(\theta
^{2}) , \label{r360}
\end{equation}
\begin{eqnarray}
\textrm{Denominator} &=&\frac{2r}{l^{2}}\left(
1-\frac{J^{2}l^{2}}{4r^{4}} \right) \sqrt{\left( 1-\frac{\beta
\theta J^{2}l^{2}}{ r^{2}(4r^{4}-J^{2}l^{2})}\right) \left( 1-\beta
\theta \left( \frac{1}{2}+ \frac{J^{2}l^{2}}{8r^{4}}\right)
\frac{4r^{2}}{4r^{4}-J^{2}l^{2}}\right) }
\nonumber \\
&&+ \textrm{Order}(\theta ^{2}) ,   \label{r29}
\end{eqnarray}
so that ignoring the higher order terms in $\theta$ Eq. (\ref{r28})
becomes

\begin{equation}
\hat{S}=\left. 4\pi \int \left( 1-\frac{\beta \theta J^{2}l^{2}}{
r^{2}(4r^{4}-J^{2}l^{2})}\right) ^{\frac{1}{2}}\left( 1-\beta \theta
(\frac{1
}{2}+\frac{J^{2}l^{2}}{8r^{4}})\frac{4r^{2}}{4r^{4}-J^{2}l^{2}}\right)
^{- \frac{1}{2}}dr \right\vert _{r=\hat{r}_{+}} . \label{r32}
\end{equation}
On integrating, this can be written as

\begin{equation}
\hat{S}=4\pi \left( \hat{r}_{+}- \frac{\beta
\theta}{\hat{r}_{+}}\right)  , \label{e3}
\end{equation}
or, on using Eq. (\ref{bt34}), we can write it as

\begin{equation}
\hat{S}=4\pi \left(  r_{+}+\frac{\beta \theta }{4} -\frac{\beta
\theta }{ r_{+}+ \beta \theta /4 }\right)  , \label{6}
\end{equation}
which is the entropy of the noncommutative BTZ black hole. Note that
the term \[4\pi \left( \frac{\beta \theta }{4}- \\ \frac{\beta
\theta }{ r_{+}+\beta \theta /4 }\right)\] is the correction
introduced by noncommutativity in the entropy of the BTZ black hole.
Again in the limit $\theta \rightarrow 0$ we recover the entropy of
the commutative BTZ black hole i.e.

\[
S=4\pi r_{+} .
\]
\newline
If we plot the mass of the noncommutative BTZ black hole in Eq.
(\ref{r19}) for different values of $\beta \theta $, we obtain the
graph as given in Fig. 1.

\FIGURE{\epsfig{file=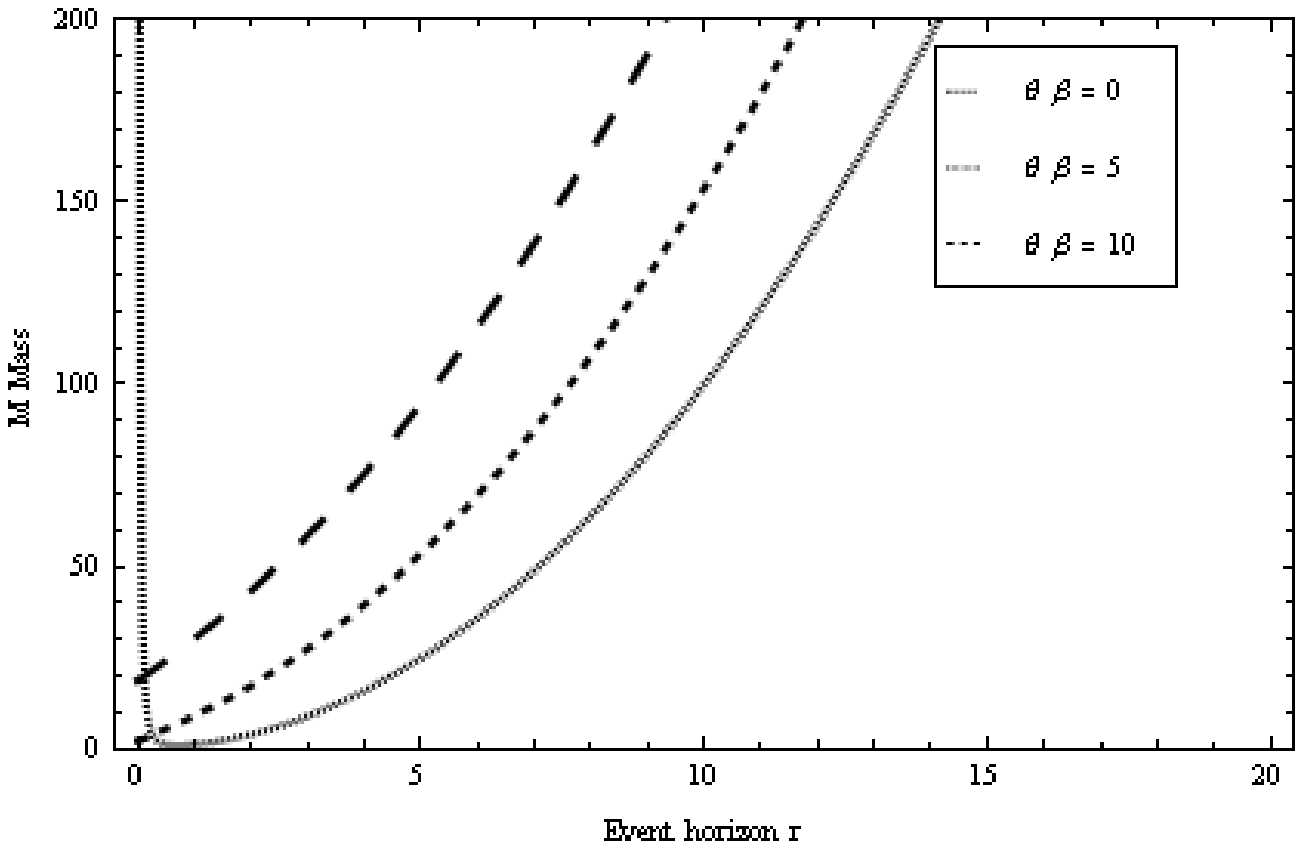,width=10cm}
        \caption[Example of figure]{Mass of the noncommutative BTZ black hole versus the event
horizon of the commutative BTZ black hole for different values of
$\beta \theta$, while setting $J=l=Q=1$.}
 \label{fig1}}

The graph shows that there is an increase in the mass of
noncommutative BTZ black hole when $\beta \theta$ is increasing from
$0$ to $10$. For the commutative case the mass becomes infinite when
we take limit $r\rightarrow 0$ but in the noncommutative case it
remains finite. Now if we plot the temperature of the noncommutative
BTZ black hole, given in Eq. (\ref{r20}) for different values of
$\beta \theta$, we get the graph as in Fig. 2.

\FIGURE{\epsfig{file=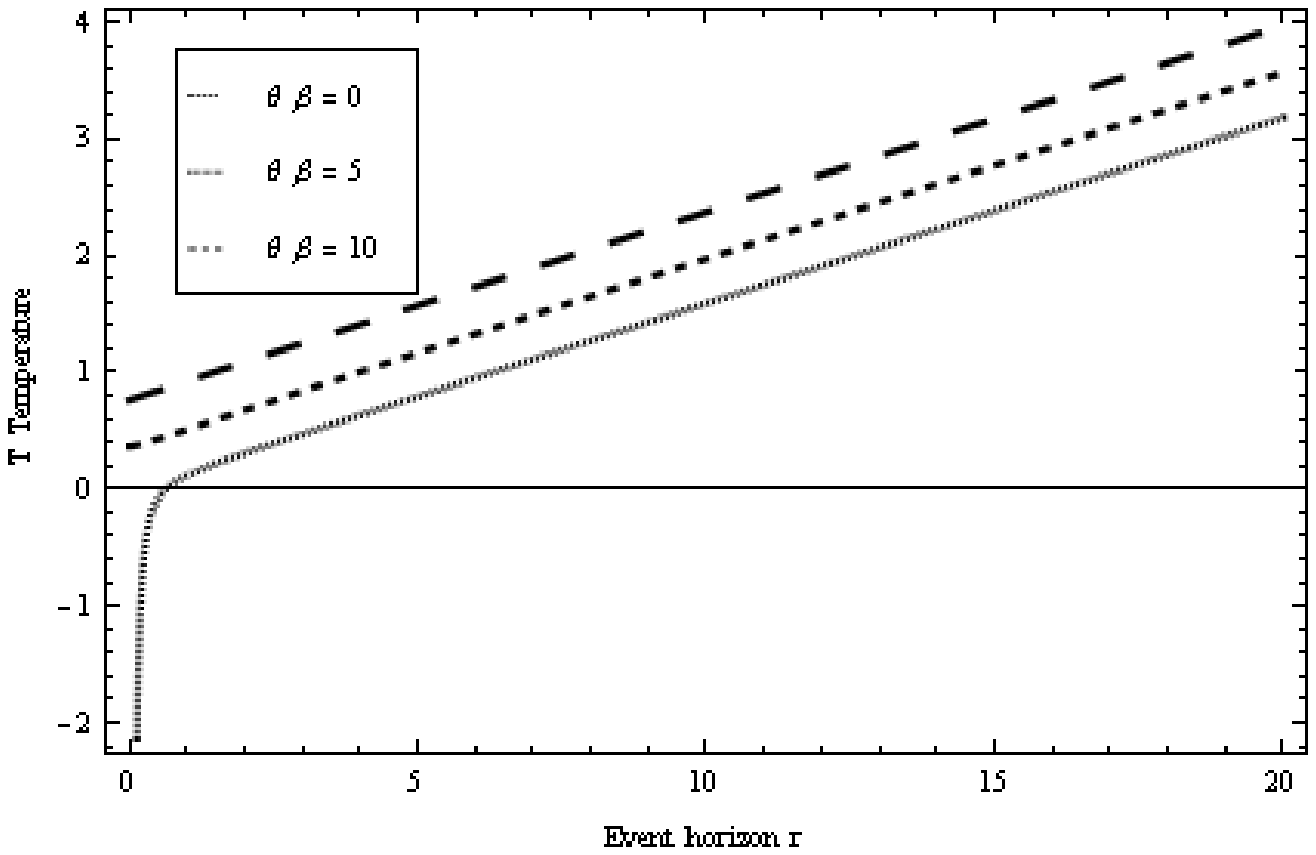,width=10cm}
        \caption[Example of figure]{Temperature of the noncommutative BTZ black hole versus
the event horizon of the commutative BTZ black hole for
different values of $\beta \theta $, while setting $J=l=Q=1$.}%
    \label{fig2}}

The entropy from Eq. (\ref{6}) for different values of $\beta
\theta$ is given in Fig. 3. The graph shows that there is an
increase in the entropy of noncommutative BTZ black hole when $\beta
\theta $ increases from 0 to 10. In the commutative case the entropy
at the origin is zero but in the noncommutative case it is nonzero.

\FIGURE{\epsfig{file=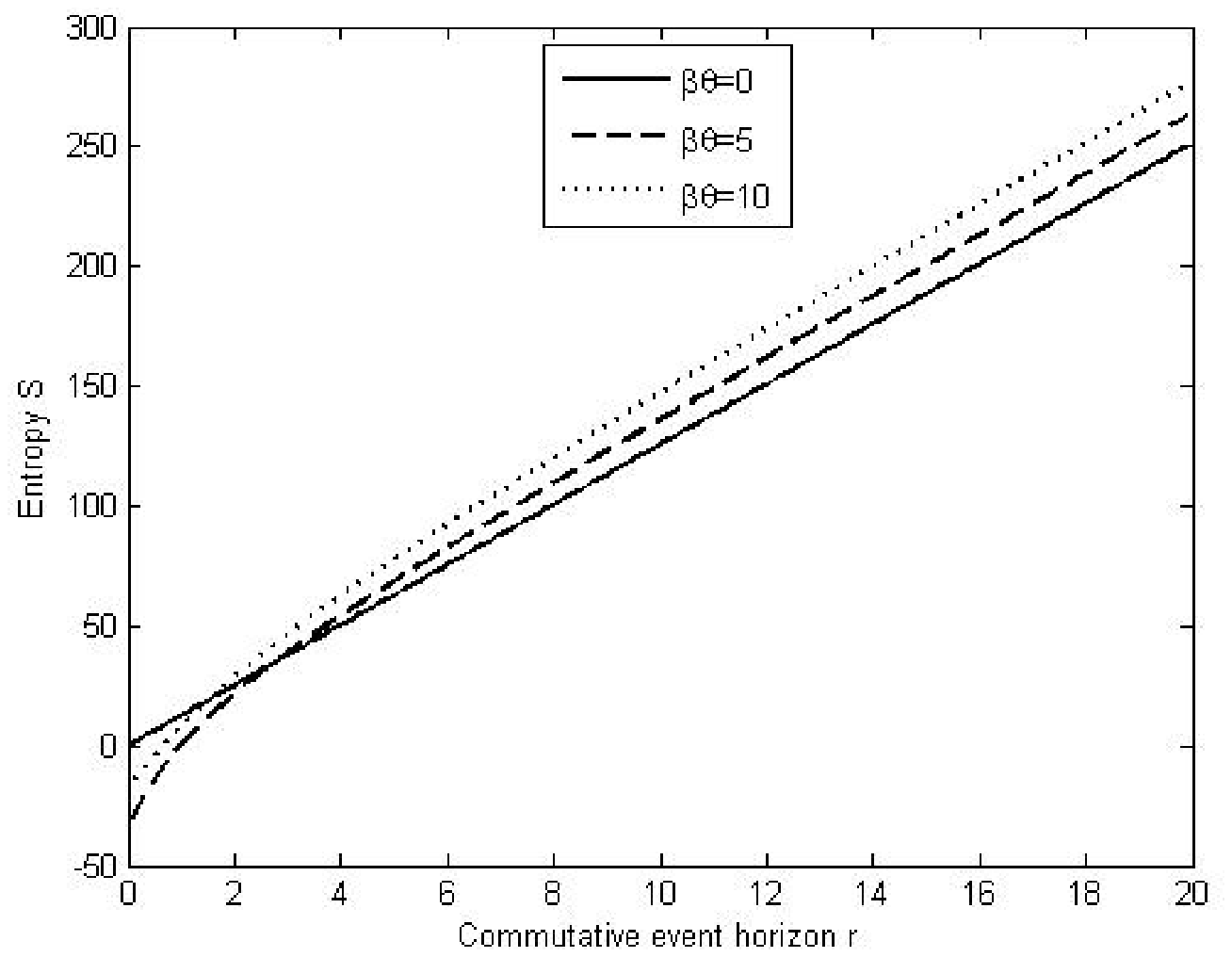,width=10cm}
        \caption[Example of figure]{Entropy of the noncommutative BTZ black hole versus the event
horizon of the commutative BTZ black hole for different values of
$\beta \theta $ while setting $J=l=Q=1$.}
    \label{fig3}}



We know that in commutative geometry the entropy of a black hole is
proportional to its area. The question is whether this area law
holds for noncommutative black holes or not. In the case of
noncommutative Schwarzschild black hole \cite{26} the law is valid
for the region $r_{h}\geq 3\sqrt{\theta}$, but for the region
$r_{h}<3\sqrt{ \theta }$ this law is not valid.

The area formula for the BTZ black hole is \cite{9}

\begin{eqnarray}
A= 16 \pi G_3 r_+ ,
\end{eqnarray}
where $G_3$ is the three dimensional Newton's gravitational
constant. Using entropy from Eq. (\ref{6}), we see that the area for
the noncommutative case takes the form

\begin{equation}
A=16 \pi \hbar G_3 \left(  r_{+}+\frac{\beta \theta }{4}
-\frac{\beta \theta }{ r_{+}+ \beta \theta/4 }\right) . \label{al}
\end{equation}

The entropy (as in Eq. (\ref{e3})) for the noncommutative BTZ black
hole is plotted against the event horizon for different values of
$\beta \theta $ in Figs. 4 and 5. We see that the area law holds for
large $r$ but as it approaches zero the law no longer remains valid.

\FIGURE{\epsfig{file=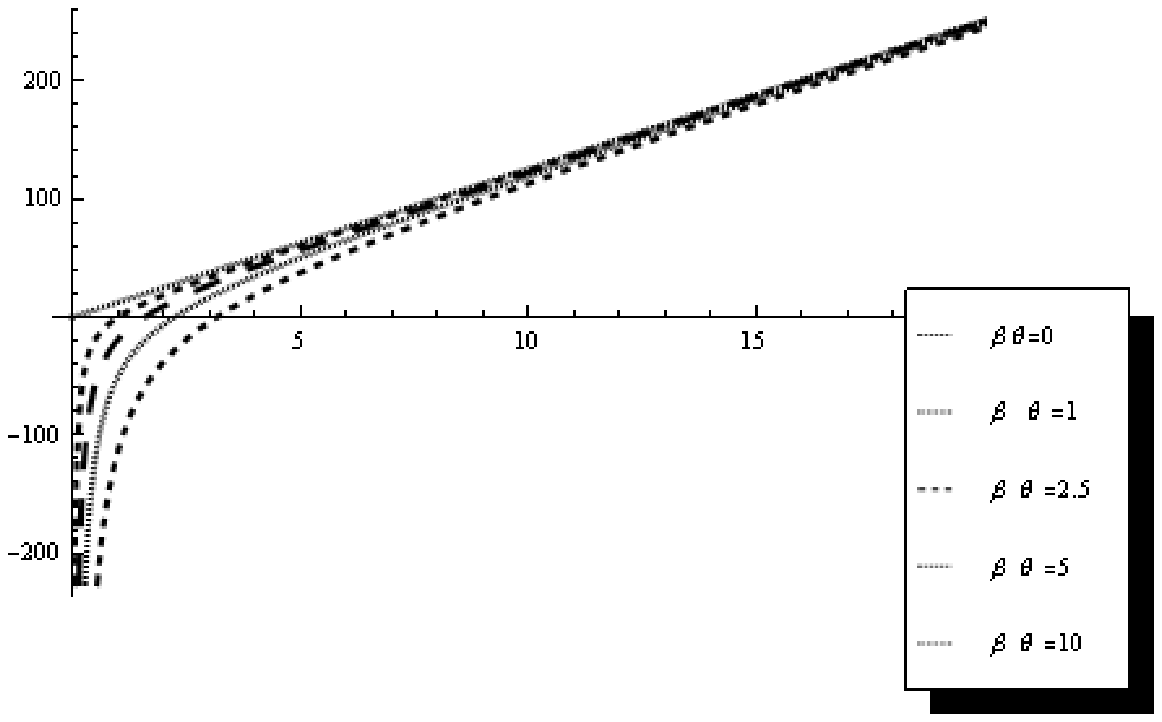,width=10cm}
        \caption[Example of figure]{Entropy of the noncommutative BTZ black hole versus the event
horizon of the noncommutative BTZ black hole}%
    \label{fig4}}

\FIGURE{\epsfig{file=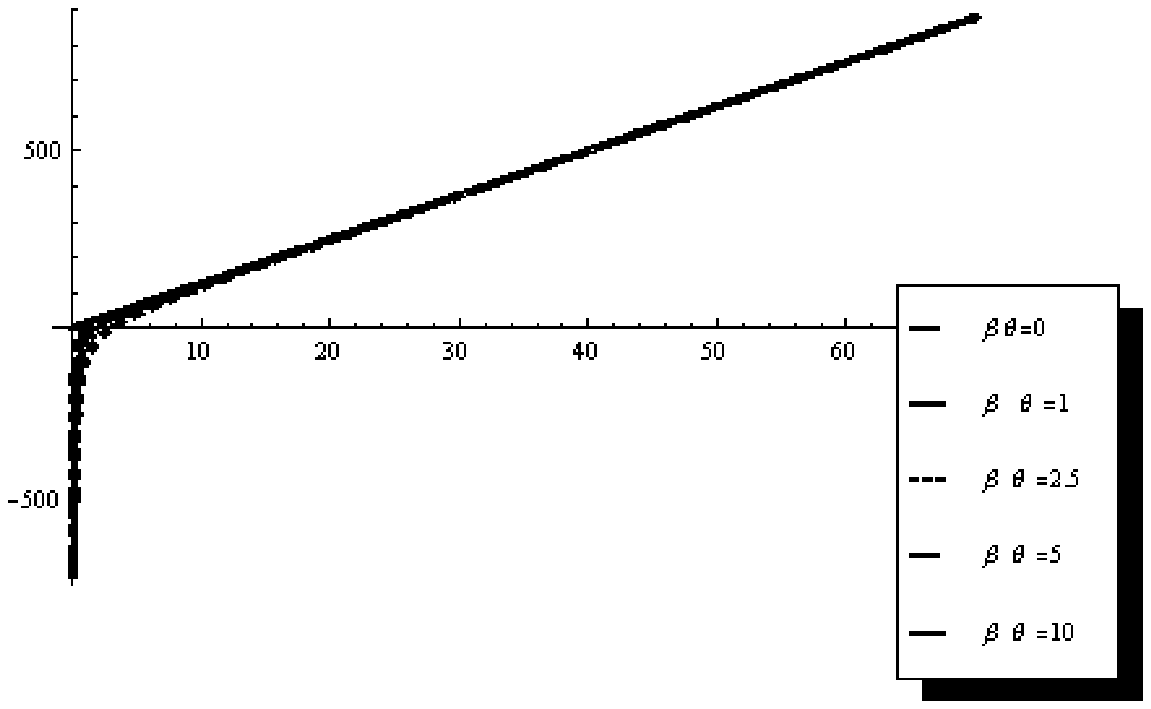,width=10cm}
        \caption[Example of figure]{Entropy of the noncommutative BTZ black hole versus the event
horizon of the noncommutative BTZ black hole (this is in fact a zoom-out of Fig. 4.)}%
    \label{fig5}}

\pagebreak

\acknowledgments

A grant from the Quaid-i-Azam University Research Fund is
acknowledged.


\begin{thebibliography}{999}

\bibitem{26} P. Nicolini, A. Smailagic and E. Spallucci, \emph{Phys. Lett. B}
\textbf{632} (2006) 547 [arXiv:gr-qc/0510112].

\bibitem{40} H-C Kim, M. Park , C. Rim and J. H. Yee, \emph{J. High Energy Phys.}
\textbf{10} (2008) 60.

\bibitem{24} P. Nicolini, \emph{Int. J. Mod. Phys. A} \textbf{24} (2009)
1229.

\bibitem{25} E. Chang-Young, D. Lee and Y. Lee, \emph{Class. Quantum Grav.} \textbf{26} (2009) 185001.

\bibitem{10} M. Banados, C. Teitelbiom and J. Zanelli, \emph{Phys. Rev. Lett.}
\textbf{69} (1992) 1849.

\bibitem{28} N. Seiberg and E. Witten, \emph{J. High Energy Phys.} \textbf{09}
(1999) 32.

\bibitem{50}R. Kerner and R. B. Mann, \emph{Phys. Rev. D} \textbf{73} (2006) 104010.

\bibitem{9} M. Akbar and K. Saifullah, \emph{Eur. Phys. J. C} \textbf{67} (2010)
205.

\end{thebibliography}
\end{document}